\documentclass{svproc}
\usepackage{float}
\usepackage{color}
\usepackage{url}

\begin{document}
\mainmatter
\title{On Generalized Langevin Dynamics and the Modelling of Global Mean Temperature.}
\titlerunning{Langevin Global Temperature Modelling ...}
\author{Nicholas W. Watkins \inst{1,2,3} \and Sandra~C.~Chapman \inst{1} \and
Aleksei Chechkin \inst{4,5} \and Ian Ford \inst{6} \and Rainer Klages \inst{7} \and David~A.~Stainforth \inst{2,1}}
\authorrunning{Nicholas Watkins et al.} 
%
%
\institute{University of Warwick, Coventry CV4 7AL, UK\\
\email{nickwatkins62@fastmail.com},\\ WWW home page:
\texttt{http://warwick.ac.uk/fac/sci/physics/research/cfsa/people/watkins/ }
\and
London School of Economics and Political Science, London WC2A 2AE, UK
\and
The Open University, Milton Keynes MK7 6AA, UK
\and
Akhiezer Institute for Theoretical Physics, Kharkov, Ukraine 
\and
Institute for Physics and Astronomy, University of Potsdam, Germany
\and
University College London, London WC1E 6BT, UK
\and
Queen Mary University of London, London E1 4NS, UK}

\maketitle

\begin{abstract}
 Climate science  employs a hierarchy of models, trading  the tractability of simplified energy balance models (EBMs)  against the detail of Global Circulation Models. Since the pioneering work of Hasselmann, stochastic EBMs have allowed treatment of climate fluctuations and noise.   However, it has recently been claimed that observations motivate heavy-tailed temporal response functions in global mean temperature to perturbations. Our complementary approach exploits the correspondence  between Hasselmann’s EBM and the original mean-reverting stochastic model in physics, Langevin’s equation of 1908.  We propose mapping a model well known in statistical mechanics, the Mori-Kubo Generalised Langevin Equation (GLE) to generalise the Hasselmann EBM. If present, long range memory then simplifies the GLE to a fractional Langevin equation (FLE).  We describe the corresponding EBMs  that map to the GLE and FLE, briefly discuss their solutions, and relate them to  Lovejoy et al’s new Fractional Energy Balance Model (FEBE).

\end{abstract}

\section{Introduction and aims of paper}
 
The importance of modelling the response of the Earth's climate to  external perturbations, whether anthropogenic or solar, and internal fluctuations, is well recognised. It has been addressed by models with many complementary levels of complexity\cite{ghil}, from general circulation models to very simplified energy balance models\cite{north}. 
Key steps included the first deterministic  EBMs, and Hasselmann's development\cite{hass,frank:hass,lemk} of these into a stochastic differential equation  with added white noise. The Hasselmann model was far from the last word in stochastic modelling for climate applications, however, and a review of progress in this area was given by the contributors to a recent book\cite{franzke}. Of particular note has been the extensive study  of multivariate stochastic models and statistical inference (e.g. \cite{pen}), and  stochastic modelling \cite{franzke2,gottwald} in climate science using the  Mori-Zwanzig formalism which provides a very general mathematical framework for the Langevin equation \cite{zwanzig,chorin}. Some of this work has benefitted from a formal equivalence between the Hasselmann model and the driven Langevin equation of statistical mechanics, and much of it has been done while still assuming that the noise terms are spectrally white. This greatly eases the   mathematical tractability by linking the problem to the most mature parts of  stochastic calculus (e.g. \cite{chorin}).  However there have long been physically-motivated arguments for making the driving noise term itself  spectrally red rather than  white. In particular Leith argued more than 25 years ago\cite{leith} that: ``unfortunately there is evidence that [a model of the white noise-driven Langevin type] would be unsatisfactory to capture some of the low frequency phenomena observed in the atmosphere. This is referred to as the {\it infrared climate problem} and appears to be caused by nonlinear interactions of the chaotic internal weather frequencies that potentially induce a ``piling-up" of extra variance at the low frequencies".

  More recently evidence has been given for long-range dependence (LRD, \cite{Beran}) in the climate system based on ``1/f" power spectra and other diagnostics\cite{franzke3}.  An alternative approach to using a ``1/f"  noise term in a stochastic EBM has been to propose non-exponential, long-tailed response functions in the time domain.  These were introduced by Rypdal\cite{rypdal} in a fractional Brownian setting. This pioneering paper was followed by further work from the Tromso group, notably \cite{fred} and \cite{rypdal2}.  More recently several papers by Lovejoy and co-workers (e.g. \cite{amador}) have demonstrated the utility of a fractional Gaussian model (StocSIPS) for climate prediction on monthly and longer time scales.  
  
  As Leith's comments implied, however, the physical arguments for extending the Hasselmann formalism are much more general than those which motivate  LRD, and there are many other well-known features of the climate system\cite{franzke,franzke3} including  periodicities and apparent spectral breaks that also need to be captured.  So  an ideal but still simple stochastic  EBM would permit a more general range of dependency structures  than either just the shortest possible range (Markovian) or longest range (LRD) behaviours, while allowing  both as limiting cases. Just such a formalism has long been available in statistical mechanics \cite{zwanzig,mori,coffey,klages1,klages2}, the Mori-Kubo generalised Langevin equation.  This is a natural extension of the Langevin equation whereby an integral over a  kernel replaces its constant damping term.

In this paper we propose  extending  the Hasselmann model by this GLE-inspired route, and show what the analogous generalised Hasselmann equation would be. We then recap how taking a power-law form for the damping kernel arrives at the fractional Langevin equation studied by researchers in topics including anomalous diffusion,  and write down a corresponding fractional Hasselmann equation.  Fractionally integrating both sides of this allows us to identify how Lovejoy et al's new \cite{lovejoy1,lovejoy} fractional energy balance equation (FEBE) relates to our  scheme. 

\section{Deterministic and stochastic energy balance models.}

The simplest energy balance models are deterministic.  They describe the (assumed linear) change $\Delta T(t)$ in the Earth's global mean temperature $T(t)$ away from its radiative equilibrium value $T_0$ (i.e. the anomaly in GMT) due to changes in the net radiative forcing $F(t)$. $\Delta T$ is multiplied by the effective heat capacity $C$ to give the change $\Delta Q$ in the heat content of the earth system, while the feedbacks in the climate system are lumped into a single parameter $\lambda$, to give a one-dimensional deterministic EBM: 
\begin{equation}
\frac{d \Delta Q}{dt} = C \frac{d \Delta T}{dt} = -\lambda \Delta T(t) + F(t)
\end{equation}

 Hasselmann's model  was a key step in improving the realism of this EBM, by incorporating delta correlated (``white") noise $\xi(t)$:
\begin{equation}
    <\xi(t)\xi(t+t')>=\sigma^2_Q \delta(t-t')
\end{equation}
giving rise to what we will refer to as the driven Hasselmann equation (DHE):
\begin{equation}
(C\frac{d}{dt} + \lambda) \Delta T(t) = 
F(t) + \xi(t)    
\end{equation}

\section{The Hasselmann-Langevin correspondence}

The DHE is the simplest possible Markovian stochastic linear  EBM, and as such,  isomorphic to the driven Langevin equation (DLE) of statistical mechanics\cite{lemons,coffey}, for the velocity of a particle executing Brownian motion: 

\begin{equation} 
(M \frac{d}{dt} + \eta) V(t) = F(t) + \xi(t)
\end{equation}
where we followed the convention of \cite{lemons} that a stochastic variable is upper case while known values of the variable are lower case. 
The correspondence is well-known in climate science, and can be expressed via  a lookup table:

\begin{table}[H]
\caption{}
\begin{tabular}{|c|c|c|c|c|c|}
\hline
  Quantity & Symbol & Units & Quantity  & Symbol &  Units  \\
  \hline
Velocity & $V(t)$ & $[m s^{-1}]$  & GMT anomaly & $\Delta T(t)$ & $[K]$   \\
Initial $V$ & $v_0$ & $[m s^{-1}]$ & Initial $\Delta T$ &   $\Delta T_0$ & $[K]$   \\
Drift $V$ & $v_d$ & $[m s^{-1}]$ & & $F/\lambda$ &  \\
Mass & $M$  &$[kg]$ & Heat capacity & $C$ & $[J K^{-1} m^{-2}]$ \\
Damping rate & $\eta/M$ &  $[s^{-1}]$ & & $\lambda/C$ & $[s^{-1}]$ \\ 
Noise strength &  $\sigma_v/M$ & $[m s^{-3/2}]$ & & $\sigma_Q/C$ & $[K s^{-1/2}]$  \\
 
\hline
\end{tabular}
\end{table}

 It should be noted that the Hasselmann model is sometimes used without a deterministic forcing term  $F(t)$ (e.g. \cite{cox}). We will refer to this as the undriven Hasselmann equation (UHE):
\begin{equation}
(C\frac{d}{dt} + \lambda) \Delta T(t) = 
 \xi(t)    
    \end{equation}
and note that it is of course in correspondence with the undriven Langevin equation
\begin{equation} 
(M \frac{d}{dt} + \eta) V(t) =  \xi(t)
\end{equation}
whose solution\cite{chorin,lemons,gard} is the well-known Ornstein-Uhlenbeck process . 

In standard texts such as \cite{gard} it is shown that in the undriven $F=0$ case, and for a solution  started at $t=0$ with velocity $v_0$, the velocity process of the ULE is
\begin{equation}
    V(t) = v_0e^{-(\eta/M)t} + \frac{1}{M} \int_0^t  e^{-(\eta/M)(t-\tau)} \xi(\tau) d\tau
\end{equation}

The equivalent solution for the temperature process in the UHE is well-known to be: 

\begin{equation}
    \Delta T(t) = \Delta T_0 e^{-(\lambda/C)t} + \frac{1}{C} \int_0^t  e^{-(\lambda/C)(t-\tau)}\xi(\tau) d\tau
\end{equation}

As noted by \cite{gard} the use of the alternative initial condition whereby the evolution starts at $t=-\infty$ means that the first term in the above can be dropped, giving 
\begin{equation}
    V(t) =   \frac{1}{M} \int_{-\infty}^t  e^{-(\eta/M)(t-\tau)} \xi(\tau) d\tau
\end{equation}
and 
\begin{equation}
    \Delta T(t) =   \frac{1}{C} \int_{-\infty}^t  e^{-(\lambda/C)(t-\tau)} \xi(\tau) d\tau
\end{equation}
respectively. The extension of the solution to the case with a  nonzero deterministic component of forcing is discussed e.g. by \cite{rypdal2}

\section{Non-Markovian models} 

We noted in the introduction, however, that in both condensed matter physics and climate science we may not always (e.g. \cite{zwanzig,hanggi}) be entitled to make the (often very good) approximation  that the noise term $\xi(t)$ is  white  and that the response  function is exponential.  Debate continues\cite{ghil,franzke2} in climate research about when a fluctuation-dissipation theorem applies (or {\it a priori} could apply), while in condensed matter applications both internal and external noises are studied\cite{klages1,klages2}, so will  not prescribe the power spectrum  of $\xi(t)$  in what follows. 
Since the 1960s one standard way to handle such cases in statistical mechanics  has been to use the full GLE rather than the Langevin equation which approximates it. In use the initial time $a$ is often be taken to be $0$ or $-\infty$ as discussed by e.g. \cite{zwanzig}.  Following the notation of \cite{lutz} but with $U'(x)=F(t)$ we have

\begin{equation}
M \frac{d}{dt}V(t) + M \int_a^t\gamma(t-\tau)V(\tau)d\tau  = 
F(t) + \xi(t)    
    \end{equation}
    
In the special case that the response is instantaneous and has no memory we have that the damping kernel  $\gamma(t)=(\eta/M)\delta(t)$ and the ordinary DLE is retrieved. 

The relevance of the GLE has been explicitly noted in climate science \cite{kondrashov} but as far as we know it has not been proposed directly as an extension of the Hasselmann EBM.
By means of the correspondence used previously  we now argue that a generalised Hasselmann equation (GHE) should be: 

\begin{equation}
C \frac{d}{dt} \Delta T(t) + C \int_a^t\gamma(t-\tau) \Delta T(\tau)d\tau  = 
F(t) + \xi(t)    
    \end{equation}
    
which would become the familiar Hasselmann equation in the limit $\gamma(t)=(\lambda/C)\delta(t)$.
    
We are not aware of a more general solution than what is available in the Mori-Zwanzig approach for the above equations for data-derived $F(t)$ of the type encountered in climate modelling.  Sophisticated numerics are now available, however
(e.g. \cite{darve}).

\section{Fractional non-Markovian models}

A special  case of the GLE which has received interest (e.g. \cite{coffey,lutz,porra,molina-garcia,kobelev}) because of its relationship to anomalous diffusion  is when the damping kernel $\gamma(t)$ is chosen to be a power law $\gamma(t) \sim t^{-\alpha}$ (e.g. \cite{lutz,porra}). The integral operator term in the GLE can then be replaced  (e.g. \cite{lutz,metzler}) by a Riemann-Liouville fractional derivative with respect to time: 
\begin{equation}
    \int_{0}^t d\tau \frac{1}{(t-\tau)^{\alpha}} \equiv \gamma_{\alpha}\frac{d^{\alpha-1}}{dt^{\alpha-1}}
\end{equation}
where the gamma function in the definition of the derivative was absorbed into $\gamma_{\alpha}$ and again we follow the notational conventions of Lutz\cite{lutz}.

In this case the GLE becomes the fractional Langevin equation (FLE) (e.g. \cite{coffey,lutz,porra,mainardi,molina-garcia}) which in Lutz' notation is:

\begin{equation}
M \frac{d}{dt}V(t) + M \gamma_{\alpha}\frac{d^{\alpha-1}}{d t^{\alpha-1}}V(t)  = 
F(t) + \xi(t)    
    \end{equation}

while the GHE that we proposed above would become a fractional Hasselmann equation (FHE):

\begin{equation}
C \frac{d}{dt}\Delta T(t)  + C\gamma_{\alpha}\frac{d^{\alpha-1}}{d t^{\alpha-1}} \Delta T(t)  = 
F(t) + \xi(t)    
\end{equation}

The exponential functions in the solutions of the Langevin equation  are known to be a special case of the solution of the FLE in terms of  Mittag-Leffler functions (e.g. \cite{coffey,lutz,porra}), and the analogous expressions would follow for the FHE (see also \cite{lovejoy}).

\section{Fractionally integrating the standard fractional non-Markovian models}

The FHE has not so far been studied in the climate literature, but a closely related equation has, Lovejoy et al's \cite{lovejoy,lovejoy1,lovejoy2} FEBE. To see this we can  integrate both sides of the FLE and FHE with respect to time, but rather than an  integral of unit order we operate with a fractional integral of order $\alpha-1$.

In the FLE case we will refer to the result as the fractionally integrated fractional Langevin equation (FIFLE):

\begin{equation}
M \frac{d^{2-\alpha}}{dt^{2-\alpha}}V(t) + M\gamma_{\alpha}V(t)  = 
\frac{d^{1-\alpha}}{dt^{1-\alpha}}(F(t) + \xi(t))
    \end{equation}
    
The allowed range of $\alpha$ is inherited from that arising from  the definition of the Riemann-Liouville fractional integral, so $\alpha >0$.
In the special case where the standard \cite{metzler} FDT for the FLE applies the noise term becomes a fractional Gaussian noise $\xi_H$ \cite{metzler,Beran} and we can also substitute for $\alpha=2-2H$ to get 

\begin{equation}
M \frac{d^{2H}}{dt^{2H}}V(t) + M\gamma_{2-2H}V(t)  = 
\frac{d^{2H-1}}{dt^{2H-1}}(F(t) + \xi_H(t))
    \end{equation}\\

    which will describe the range $1/2<H<1$, i.e. when an fBm which integrates the  fractional Gaussian noise term $\xi_H$ would be superdiffusive.

The corresponding fractionally integrated fractional Hasselmann equation (FIFHE) would then be 
    
    \begin{equation}
C \frac{d^{2-\alpha}}{dt^{2-\alpha}}\Delta T(t)  + C\gamma_{\alpha}\Delta T(t)  = 
\frac{d^{1-\alpha}}{dt^{1-\alpha}}(F(t) + \xi(t))
    \end{equation}
    
and if we posit that an FDT having the same functional form as the FLE's applies then we have
\begin{equation}
C \frac{d^{2H}}{dt^{2H}}\Delta T(t) + C\gamma_{2-2H}\Delta T(t)  = 
\frac{d^{2H-1}}{dt^{2H-1}}(F(t) + \xi_H(t))
    \end{equation}\\
which again applies to the range $1/2<H<1$. These two versions of the FIFHE are already of a similar form to Lovejoy et al's FEBE \cite{lovejoy,lovejoy2}: 
\begin{equation}
 (\tau^{H_L}\frac{d^{H_L}}{dt^{H_L}} + 1)\Delta T(t) 
= \frac{\tau}{C} 
(F(t) + \xi(t))
\end{equation}\\

where $H_L$ (denoted by ``H" in their papers)  is not the usual Hurst exponent $H$ of  Mandelbrot's fBm and fGn  (e.g. \cite{Beran,watkins}), which we used above. 

To further pin down the relationship between the FIFHE and  FEBE  we can  identify $H_L$ as $1+d$ where $d$ is the  memory parameter  used in much of the mathematical  and statistical  literature (e.g. \cite{Beran}) on stochastic processes which exhibit LRD including $ARFIMA(p,d,q)$.  Thus $H_L=1,H=1/2,d=0$  is the limit corresponding to Hasselmann's original white noise driven, exponentially relaxing equation, and the range $1/2<H_L<3/2$ is that corresponding to fBm. 

If we then identify $1+d$ with our $2-\alpha$, define a time scale $\tau$ via $\tau^{1+d}  =\gamma^{-1}_{\alpha}$, and multiply across by $\tau^{1+d}/C$ we see that   the FIFHE becomes:\\

\begin{equation}
 (\tau^{1+d}\frac{d^{1+d}}{dt^{1+d}} + 1)\Delta T(t) 
= \frac{\tau^{1+d}}{C} \frac{d^{d}}{dt^{d}} (F(t) + \xi(t))
\end{equation}\\

The left hand side of the FIFHE is then seen to be isomorphic to  that of the FEBE. The right hand side differs from the FEBE, being a   fractional derivative of the sum of the forcing term and a noise, the details of which will depend on those of $\xi$ itself. We stress that this result is independent of whether or not an FDT has yet been assumed.

In the special case where an FDT of the same form as that of the FLE \cite{metzler} applies we use  $\alpha=2-2H$ to give $d=2H-1$ and then, as before, define a time scale $\tau$ by $\tau^{2H}=\gamma^{-1}_{2-2H}$, and multiply across by $\tau^{2H}/C$ to give:

\begin{equation}
 (\tau^{2H}\frac{d^{2H}}{dt^{2H}} + 1)\Delta T(t) 
= \frac{\tau^{2H}}{C} \frac{d^{2H-1}}{dt^{2H-1}} (F(t) + \xi_H(t))
\end{equation}\\
 
 which, as we remarked in \cite{onlyurl1} applies in the range $H<1/2<1$.

\section{Conclusions}
This paper is a first report on our project \cite{onlyurl1,onlyurl2} to use the well-known Mori-Kubo generalised Langevin equation of statistical mechanics to seek a flexible framework in which the Hasselmann equation and  our FIFHE are respectively the short range and long range correlated limits.  We intend to further explore the properties of the FHE and FIFHE and their relation to the FEBE in future work, as well as those of the more general GHE for climate.  \\

\noindent {\bf Acknowledgements} We acknowledge valuable interactions with and encouragement from Raphael Calel,  Michael Ghil, Valerio Lembo, Shaun Lovejoy, Valerio Lucarini, Nick Moloney, Cecile Penland, Kristoffer Rypdal and Martin Rypdal. NWW would particularly like to thank the organisers of the Exeter  University workshop on Emergent Constraints on 23rd November 2020,  questions at which from Peter Cox and Jan Sieber  spurred the production of the revised version of this preprint. 
He also acknowledges support from the Max Planck Society, London Mathematical Laboratory, KLIMAFORSK project number 229754 and ONR grant NICOP – N62909-15-1-N143. NWW and SCC acknowledge the hospitality of Holger Kantz at MPIPKS, Ralf Metzler at the University of Potsdam,  Juergen Kurths at PIK and Sid Redner at SFI during the development of this work.  SCC  acknowledges support from STFC grant ST/T000252/1 and AFOSR grant FA9550-17-1-0054. DAS acknowledges support from the Grantham Research Institute on Climate Change and the Environment, at the London School of Economics, and the ESRC Centre for Climate Change Economics and Policy (CCCEP) (ref. ES/R009708/1).
%
%

\end{document}